\documentclass[12pt]{article}

\usepackage{epsfig}
\usepackage{amsmath}
\usepackage{soul,color}
\usepackage{bm}
\usepackage{epsfig}
\usepackage{natbib}
\usepackage{hyperref}
\usepackage[]{graphicx}\usepackage[]{color}
\makeatletter
\def\maxwidth{ %
  \ifdim\Gin@nat@width>\linewidth
    \linewidth
  \else
    \Gin@nat@width
  \fi
}
\makeatother

\definecolor{fgcolor}{rgb}{0.345, 0.345, 0.345}
\newcommand{\hlnum}[1]{\textcolor[rgb]{0.686,0.059,0.569}{#1}}%
\newcommand{\hlstr}[1]{\textcolor[rgb]{0.192,0.494,0.8}{#1}}%
\newcommand{\hlopt}[1]{\textcolor[rgb]{0,0,0}{#1}}%
\newcommand{\hlstd}[1]{\textcolor[rgb]{0.345,0.345,0.345}{#1}}%
\newcommand{\hlkwb}[1]{\textcolor[rgb]{0.69,0.353,0.396}{#1}}%
\newcommand{\hlkwc}[1]{\textcolor[rgb]{0.333,0.667,0.333}{#1}}%
\newcommand{\hlkwd}[1]{\textcolor[rgb]{0.737,0.353,0.396}{\textbf{#1}}}%

\usepackage{framed}
\makeatletter
\newenvironment{kframe}{%
 \def\at@end@of@kframe{}%
 \ifinner\ifhmode%
  \def\at@end@of@kframe{\end{minipage}}%
  \begin{minipage}{\columnwidth}%
 \fi\fi%
 \def\FrameCommand##1{\hskip\@totalleftmargin \hskip-\fboxsep
 \colorbox{shadecolor}{##1}\hskip-\fboxsep
     \hskip-\linewidth \hskip-\@totalleftmargin \hskip\columnwidth}%
 \MakeFramed {\advance\hsize-\width
   \@totalleftmargin\z@ \linewidth\hsize
   \@setminipage}}%
 {\par\unskip\endMakeFramed%
 \at@end@of@kframe}
\makeatother

\definecolor{shadecolor}{rgb}{.97, .97, .97}
\definecolor{messagecolor}{rgb}{0, 0, 0}
\definecolor{warningcolor}{rgb}{1, 0, 1}
\definecolor{errorcolor}{rgb}{1, 0, 0}
\newenvironment{knitrout}{}{} 

\usepackage{alltt}

\newcommand{\tr}{^{\prime}}

\def\b#1{\mbox{\boldmath $#1$}}    
\def\bl#1{\mbox{\footnotesize \boldmath {$#1$}}} 

\renewcommand{\th}{\theta}
\newcommand{\al}{\alpha}

\newcommand{\bu}{\bar{u}}

\usepackage{epsfig}
\usepackage{pstricks}
\usepackage{soul,color}

\newcommand{\pkg}[1]{{\normalfont\fontseries{b}\selectfont #1}}
\let\proglang=\textsf
\let\code=\textsf

\def\baselinestretch{1.5}
\usepackage{bm}
\usepackage{epsfig}

\begin{document}

\title{\pkg{LMest}: an \proglang{R} package for latent Markov models for categorical longitudinal data}

\author{Francesco Bartolucci\\{\footnotesize University of Perugia}  \and Alessio
  Farcomeni\\{\footnotesize Sapienza - University of Rome} \and Silvia
  Pandolfi\\{\footnotesize University of Perugia} \and Fulvia
  Pennoni\\{\footnotesize University of Milano-Bicocca}}


\maketitle

\def\baselinestretch{1.2}

\begin{abstract}
Latent Markov (LM) models represent an important class of models for the analysis of longitudinal data \citep{bart:farc:penn:13}, especially when response variables are categorical.
These models have a great potential of application for the analysis of social, medical, and behavioral data as well as in other disciplines. 
We propose the \proglang{R} package \pkg{LMest}, 
which is tailored to deal with these types of model. 
In particular, we consider a general framework for extended LM models by including individual covariates and by formulating a mixed approach to take into account additional dependence structures in the data.  
Such extensions lead to a very flexible class of models, which allows us to fit different types of longitudinal data.  
Model parameters are estimated through the expectation-maximization algorithm, based on the forward-backward recursions, which is implemented in the main functions of the package.
The package also allows us to perform local and global decoding and  to obtain standard errors for the parameter estimates.
We illustrate its use and the most important features on the basis of examples involving applications in health and criminology.

\end{abstract}\newpage

\def\baselinestretch{1.1}

%



\section{Introduction}


In this paper we illustrate the \proglang{R} package
\pkg{LMest} V2.0, which replaces version V1.0, 
available from \url{http://CRAN.R-project.org/package=LMest}.
The new version extends V1.0 in several ways and it is aimed at providing a collection of tools
that can be used to estimate the class of Latent Markov (LM) models
for longitudinal categorical data.
The package can be seen as a companion to the book \cite{bart:farc:penn:13} that is focused on these models; we will closely refer to this book for the theoretical and some practical details.
Additional insights are given in the discussion paper by \cite{bart:farc:penn:14}. 

LM models are designed to deal with univariate and multivariate longitudinal data based on the repeated observation of a panel of subjects across time. 
More in detail, LM models are specially tailored to study the evolution of an individual characteristic of interest that is not directly observable.
This characteristic is represented by a latent process following a Markov chain as in a Hidden Markov (HM) model \citep{zucc:macd:09}. 
These models also allow us to account for time-varying unobserved heterogeneity in addition to the effect of observable covariates on the response variables.

The initial LM formulation introduced by \cite{wigg:73}, see also \cite{wigg:55}, has been developed in several directions and in connection with applications in many fields, such as psychology, sociology, economics, and medicine. 
In particular, the basic LM model, relying on a homogenous Markov chain of first order, 
has several extensions based on parameterizations that allow us to include hypotheses and constraints of interest. 
The most important extension is for the inclusion of individual covariates that may affect the distribution of the latent process \citep{verm:lang:bock:99,bart:penn:fran:07} or the conditional distribution of the response variables given this process \citep{bart:farc:09}.
LM models may also be formulated to take into account additional dependence structures in the data. 
In particular, the basic LM model has been generalized to the mixed LM approach by \cite{vand:lang:90} and to the LM model with random effects \citep{Altman:07,maru:11, bart:penn:vitt:11}.
These formulations are related to extended models in which
the parameters are allowed to vary in different latent subpopulations; 
for a complete review see \cite{bart:farc:penn:13} and \cite{bart:farc:penn:14}.

As already mentioned, LM models are closely related to HM models \citep{zucc:macd:09}, which are specific for time-series data.
LM models are formulated quite similarly, but they are tailored for longitudinal data where many subjects are observed only at few occasions, typically no more than ten.
Most importantly, differently from LM models, extensions of HM models to include covariates or for complex data structures are rarely applied because these structures are typical of longitudinal studies.  

In \proglang{R} there are some packages that can handle models that are related to LM models.  
In particular, HM models can be fit through packages
\pkg{HMM} \citep{himm:10}, \pkg{HiddenMarkov}
\citep{hart:12}, and \pkg{depmixS4} \citep{viss:spee:10}. 
The last one is the most closely related to \pkg{LMest} as it is tailored to deal with HM models based on a generalized linear formulation, which can be specified conditionally on individual covariates.
\pkg{depmixS4} is nonetheless designed for repeated measurements on a single unit, as in a time-series context.  
Packages \pkg{mhsmm} \citep{conn:hojs:11} and \pkg{hsmm} \citep{bull:13} deal with hidden semi-Markov models, which are HM models in which a separate assumptions are specified for persistency in a latent state.
Finally, package \pkg{msm} \citep{jack:11} can deal with HM and related models evolving in continuous time. 
A commercial software which can be used to fit certain LM models is \proglang{Latent Gold} \citep{verm:magi:03}. 
Additionally, certain \proglang{Matlab} toolboxes are also available for this aim; see for example the HMM toolbox proposed by \cite{murphy_web}. 

The distinguishing features of the \pkg{LMest} package with respect to the ones above are the following:
\begin{itemize}
\item it is primarily designed to work with longitudinal data, that is, with (even many) i.i.d. replicates of (usually short) sequences;
\item it can deal with univariate and multivariate categorical outcomes;
\item  standard errors for the parameter estimates are obtained by exact computation or through reliable approximations of the observed information matrix; 
\item additional random effects can be used to build mixed LM models;
\item individual covariates are included through efficient parameterizations of either the measurement or the latent model; 
\item computationally efficient algorithms are implemented for predictions; 
\item the package is flexible and it minimizes the use of time resources by relying on certain \proglang{Fortran} routines. 
\end{itemize}

The paper is organized as follows. Section \ref{sec:lm} briefly outlines the general formulation of LM models
and their maximum likelihood estimation, together with a brief description of the basic LM model. Section \ref{sec:measurement_cov} describes the use of the  \pkg{LMest}
package to estimate LM models with individual covariates included in the measurement model, while in Section \ref{sec:covlat} we deal with individual covariates affecting the distribution of the latent process.
Both formulations are illustrated through examples based
on real data about self-reported health status.
In Section \ref{sec_mixed} we introduce the mixed LM models and we describe the use of the \proglang{R} function aimed at estimating this model formulation.
Here, the illustrative example is based on data in the field of criminology. 
Finally, Section \ref{sec:conclusions} contains the main conclusions.
\section{Latent Markov models for longitudinal data}\label{sec:lm}
In the following we provide a brief review of the statistical methodology on which the LM models are based, trying to reduce the mathematical details as much as possible.
The illustration closely follows the recent paper by \cite{bart:farc:penn:14}.
We focus in particular on the maximum likelihood estimation of these models on the basis of the Expectation-Maximization (EM) algorithm \citep{demp:lair:rubi:77}.
We also deal with more advanced topics which are important for the correct use of the \pkg{LMest} package, such as  the selection of the number of latent states and local and global decoding \citep{vite:67,juan:rabi:91}.

\subsection{The general latent Markov model formulation}\label{sec:LMgeneral}
Consider the multivariate case where we observe a vector $\b Y^{(t)}$ of $r$ categorical response variables, $Y_{j}^{(t)}$, with $c_j$ categories, labeled from 0 to $c_j-1$, $j=1,\ldots,r$, which are available at the $t$-th
time occasion, $t=1,\ldots,T$.
Also let $\tilde{\b Y}$ be the vector obtained by stacking $\b Y^{(t)}$ for $t=1,\ldots,T$; this vector has then $rT$ elements.
Obviously, in the univariate case we have a single response variable $Y^{(t)}$ for each time occasion, and $\tilde{\b Y}$ is made by $T$ elements.
When available, we also denote by $\b X^{(t)}$ the vector of individual covariates
at the $t$-th time occasion and by $\tilde{\b X}$ the vector of all the individual
covariates which is obtained by stacking the vectors $\b X^{(t)}$, for $t=1,\ldots,T$.

The general formulation of the model assumes the existence of a
latent process, denoted by $\b U =(U^{(1)},\ldots,U^{(T)})$,
which affects the distribution of the response variables. Such a process is assumed
to follow a first-order Markov chain with state space
$\{1,\ldots,k\}$, where $k$ is the number of latent states. Under the
{\em local independence} assumption, the response vectors $\b Y^{(1)},\ldots,\b Y^{(T)}$ are
assumed to be conditionally independent given the latent process $\b U$.
Moreover, the elements $Y_{j}^{(t)}$ within $\b Y^{(t)}$, $t=1,\ldots,T$, are conditionally independent given $U^{(t)}$.
This assumption leads to a strong simplification of the model, but it can be relaxed by allowing serial dependence through the inclusion of the lagged response
variables among covariates \citep[see][among others]{bart:farc:09}.

Parameters of the measurement model are the conditional response probabilities
\[
\phi_{jy|u\bl x}^{(t)}=p(Y^{(t)}_{j}=y|U^{(t)}=u, \b X^{(t)}=\b x), \quad
j=1,\ldots,r, \;\;y=0,\ldots,c_j-1,
\]
which reduce to 
\[
\phi_{y|u\bl x}^{(t)}=p(Y^{(t)}=y|U^{(t)}=u, \b X^{(t)}=\b x), \quad
y=0,\ldots,c-1,\]
in the univariate case, where $t=1,\ldots,T$ and $u=1,\ldots,k$.

Parameters of the latent process are the initial probabilities
\[
\pi_{u|\bl x}=p(U^{(1)}=u|\b X^{(1)}=\b x),\quad u=1,\ldots,k,\]
and the transition probabilities
\[\pi_{u|\bu\bl x}^{(t)}=p(U^{(t)}=u|U^{(t-1)}=\bu,\b
X^{(t)}=\b x), \quad t=2,\ldots,T, \;\bu,u=1,\ldots,k.\]

In the expressions above, $\b x$ denotes a realization of $\b X^{(t)}$, $y$ denotes a realization
of $Y_j^{(t)}$, $u$ denotes a realization of $U^{(t)}$, and $\bu$ denotes a realization 
of $U^{(t-1)}$. 

On the basis of the above parameters, the conditional distribution of $\b U$   
given $\tilde{\b X}$ may be expressed as
\[
p(\b U=\b
u|\tilde{\b X}=\tilde{\b x})=\pi_{u^{(1)}|\bl x^{(1)}}\prod_{t>1}\pi_{u^{(t)}|u^{(t-1)}\bl x^{(t)}},
\]
where $\b u = (u^{(1)},\ldots,u^{(T)})$ and $\tilde{\b x}$ denotes a realization of the vector of all
response variables $\tilde{\b X}$. Moreover, the conditional
distribution of $\tilde{\b Y}$ given $\b U $ and $\tilde{\b X}$ may be expressed as
\[
p(\tilde{\b Y}=\tilde{\b y}|\b U=\b u, \tilde{\b X}=\tilde{\b x}) = \prod_t
\phi_{\bl y^{(t)}|u^{(t)}\bl x^{(t)}}^{(t)},
\]
where, in general, we define $\phi_{\bl y|u\bl x}^{(t)} = p(\b Y^{(t)}=\b y|U^{(t)}=u,\b X^{(t)}=\b x) $ and, due to the assumption of local independence, we have 
\[
\phi_{\bl y|u\bl x}^{(t)}=
\prod_j \phi_{jy_j|u\bl x}^{(t)}. 
\]
Note that in the above expressions $\tilde{\b y}$ is a realization of $\tilde{\b Y}$ made by the subvectors $\b y^{(t)}=(y_1^{(t)},\ldots,y_r^{(t)})$
whereas $\b y$ is a realization of $\b Y^{(t)}$ with elements $y_j$, $j=1,\ldots,r$.

When we deal with the inclusion of individual covariates,  the {\em manifest distribution} of the response
variables corresponds to the conditional distribution of $\tilde{\b Y}$ given $\tilde{\b X}$, defined as
\begin{eqnarray}
p(\tilde{\b y}|\tilde{\b x}) = p(\tilde{\b Y}=\tilde{\b y}|\tilde{\b X}=\tilde{\b x})& = & \sum_{\bl u} \pi_{u^{(1)}|\bl x^{(1)}}\pi_{u^{(2)}|u^{(1)}\bl x^{(2)}}^{(2)}\cdots
\pi_{u^{(T)}|u^{(T-1)}\bl x^{(T)}}^{(T)}\nonumber\\
&&\times \phi_{\bl y^{(1)}|u^{(1)}\bl x^{(1)}}^{(1)}\cdots
\phi_{\bl y^{(T)}|u^{(T)}\bl x^{(T)}}^{(T)}. \label{eq:dist3}
\end{eqnarray}

In the basic version of the model, all assumptions are not formulated conditionally on individual covariates. 
Moreover, when such covariates are available, we suggest to avoid that they simultaneously affect both the distribution of the latent process and the conditional distribution of the response variables given this process.
In fact, the two formulations have different interpretations, as explained in more 
detail in the following, and the resulting model would be of difficult interpretation and estimation.

Finally, it is important to note that computing $p(\tilde{\b y}|\tilde{\b x})$, or $p(\tilde{\b y})$ in the basic version, involves a sum extended to
all the possible $k^T$ configurations of the vector $\b u$;  this
typically requires a considerable computational effort.
In order to efficiently compute such a probability we can use a forward recursion \citep{baum:et:al:70}; see \cite{bart:farc:penn:13} for details.
\subsection{Maximum likelihood estimation}\label{sec:EM}
Given a sample of $n$ independent units that provide the response vectors $\tilde{\b y}_1, \ldots, \tilde{\b y}_n$, and given the corresponding vectors of covariates $\tilde{\b x}_1,\ldots,\tilde{\b x}_n$, the model log-likelihood assumes the following expression
\[
\ell(\b\th) = \sum_{i=1}^n\log p(\tilde{\b y}_i|\tilde{\b x}_i ).
\]
Note that in this case each vector $\tilde{\b y}_i$ is a realization of $\tilde{\b Y}_i$ which, in the multivariate case, is made up of the 
subvectors $\b y_i^{(t)}$, $t=1,\ldots,T$, that, in turn, have elements $y_{ij}^{(t)}$, $j=1,\ldots,r$. 
Moreover, $\tilde{\b x}_i$ may be decomposed into the time-specific subvectors of covariates
$\b x_i^{(1)},\ldots,\b x_i^{(T)}$.
Also note that $p(\tilde{\b y}_i|\tilde{\b x}_i )$ corresponds to the manifest probability of the responses provided by subject $i$, given the covariates; see equation (\ref{eq:dist3}).  
Moreover, $\b\th$ is the vector of all free parameters affecting $p(\tilde{\b y}_i|\tilde{\b x}_i)$.

The above log-likelihood function can be maximized by the EM algorithm \citep{baum:et:al:70,demp:lair:rubi:77}, as described in the following section.

\subsubsection{Expectation-Maximization algorithm}\label{Em}
The EM algorithm is based on the complete data log-likelihood that, with multivariate categorical data, has the following expression
\begin{eqnarray}
\hspace*{-1.5cm}\ell^*(\b\th) &=& \sum_{j=1}^r \sum_{t=1}^T\sum_{u=1}^k\sum_{\bl x} \sum_{y=0}^{c_j-1} a_{ju\bl x y}^{(t)}\log \phi_{jy|u\bl x}^{(t)}+\sum_{u=1}^k \sum_{\bl x} b_{u\bl x}^{(1)}\log \pi_{u|\bl x}
\nonumber\\
&& +
\sum_{t=2}^T\sum_{\bu=1}^k\sum_{u=1}^k\sum_{\bl x} b_{\bu u \bl x}^{(t)}\log \pi_{u|\bu\bl x}^{(t)},\label{eq:comp_lk_cov}   
\end{eqnarray}
where, with reference to occasion $t$ and covariate configuration $\b x$,  $a_{ju\bl x y}^{(t)}$ is the number of subjects
that are in the latent state $u$ and provide response $y$ to response variable $j$,
$b_{u\bl x}^{(t)}$ is the frequency for latent state $u$, and  $b_{i\bu u\bl x}^{(t)}$ is the number of transitions
 from state $\bu$ to state $u$. 

The EM algorithm alternates the following two steps until convergence:
\begin{itemize}
\item {\bf E-step}: it consists of computing the posterior (given the observed data) expected value of each indicator variable involved in (\ref{eq:comp_lk_cov}) by suitable forward-backward recursions \citep{baum:et:al:70}; these expected values are denoted by
$\hat{a}_{ju\bl x y}^{(t)}$, $\hat{b}_{u\bl x}^{(t)}$, and  $\hat{b}_{\bu u\bl x}^{(t)}$;
\item {\bf M-step}: it consists of maximizing the complete data log-likelihood expressed as in (\ref{eq:comp_lk_cov}), with each indicator variable substituted by the corresponding expected value.
How to maximize this function depends on the specific formulation of the model and,
in particular, on whether the covariates are included in the
measurement model or in the latent model.
\end{itemize}

We refer the reader to \cite{bart:farc:penn:13} for a detailed description of these steps, which are implemented in suitable functions of the \pkg{LMest} package.

The EM algorithm
could converge to a mode of the log-likelihood which does not correspond to the global maximum.
In this regard, we suggest to combine a deterministic and a random starting rule for the EM algorithm, so to take as final estimate the one corresponding to the highest log-likelihood obtained at convergence of the algorithm; the corresponding estimate is denoted by $\hat{\b\th}$.
This may prevent problems due to the multimodality of this function.
In particular, the random initialization is based on suitably normalized random numbers drawn from a uniform distribution from 0 to 1 for the initial and transition probabilities of the Markov chain and for the conditional response probabilities.
The convergence of the EM algorithm is checked on the basis of the relative log-likelihood difference, that is,
\begin{equation}
(\ell (\b \th)^{(s)} -\ell (\b \th)^{(s-1)})/|\ell(\b\th)^{(s)}| < \epsilon,\label{eq:tol1}
\end{equation}
where  $\b \th^{(s)}$ is the parameter estimate obtained at the end of the $s$-th M-step and
$\epsilon$ is a predefined tolerance level which can be set by the users. 

\subsubsection{Standard errors}
After the model is estimated, standard errors may be obtained on the basis of the observed information matrix, denoted by $\b J(\hat{\b\th})$, by standard rules.
In particular, they are computed through the squared root of the diagonal elements of the inverse of this matrix.
The \pkg{LMest} package computes this matrix, and then provides the standard errors, by using either the numerical method proposed by \cite{bart:farc:09} or the exact computation method proposed by \cite{bart:farc:14}, according to the model of interest.  

According to the approximated method, $\b J(\hat{\b\theta})$ is obtained as minus the numerical derivative of the score vector $\b s(\hat{\b\theta})$ at convergence.
The score vector, in turn, is obtained as the first derivative of the expected value of the complete data log-likelihood, which is based on the expected frequencies $\hat{a}_{ju\bl x y}^{(t)}$, $\hat{b}_{u\bl x}^{(t)}$, and $\hat{b}_{\bu u\bl x}^{(t)}$ computed with the final parameter estimates $\hat{\b\th}$; for details see \cite{bart:farc:penn:13} and \cite{penn:14}.

The exact computation of $\b J(\hat{\b\theta})$ is instead based on the Oakes' identity \citep{oake:99}.  
This method uses the complete data information matrix, produced by the EM algorithm, and a correction matrix computed on the basis of the first derivative of the posterior probabilities obtained from the backward-forward recursions.

For the basic LM model and for the model with individual covariates affecting the distribution of the latent process, the \pkg{LMest} package also provides functions aimed at performing a parametric bootstrap \citep{davison:hynkley:1997} in order to obtain standard errors.  

\subsection{Criteria for selecting the number of latent states}
In certain applications, the number of latent states, $k$, can be {\em a priori} defined, as in the univariate case in which the number of latent states may be fixed  equal to the number of response categories.
Otherwise, the following criteria are typically used to select the number of latent states:  Akaike information criterion (AIC) of \cite{aka:73} and Bayesian information criterion (BIC) of \cite{sch:78}.
They are based on the indices 
\begin{eqnarray*}
AIC&=& -2 \hat{\ell}  + 2 \; {\rm np},\\
BIC&=& -2  \hat{\ell} + \log(n)\;   {\rm np},
\end{eqnarray*}
where $\hat{\ell}$ denotes  the maximum of the log-likelihood of the model of interest and ${\rm np}$ denotes the number of its free parameters.
 
According to the above criteria, the optimal number of latent states is the one corresponding to the minimum value of the above indices; this model represents the best compromise between goodness-of-fit and complexity.  
However, other criteria may be used, such as those taking into account the quality of the classification; for reviews see \cite{bacc:pand:penn:14} and \cite{bart:bacc:penn:14}.   
\subsection{Local and global decoding}
The \pkg{LMest} package allows us to perform decoding, that is, prediction of the sequence of the latent states for a certain sample unit on the basis of the data observed for this unit.

In particular, the EM algorithm directly provides the estimated posterior probabilities  
of $U^{(t)}$, that is, $p(U^{(t)}=u| \tilde{\b X}=\tilde{\b x},\tilde{\b Y}=\tilde{\b y})$, for $t=1,\ldots,T$,  $u=1,\ldots,k$, and for every covariate and response configuration $(\tilde{\b x},\tilde{\b y})$ observed at least once.
These probabilities can be directly maximized to obtain a prediction of the latent state of each subject at each time occasion $t$; this is the so-called {\it local decoding}. 
It shall be noted that local decoding minimizes the classification error at each time occasion, but may yield sub-optimal or even inconsistent predicted sequences of $U^{(1)},\ldots,U^{(T)}$. 

In order to track the latent state of a subject over time, the most {\em a posteriori} likely sequence of states must be obtained, through the so-called {\it global decoding}, which is based on an adaptation of the \cite{vite:67} algorithm; see also \cite{juan:rabi:91}. 
The algorithm proceeds through a forward-backward recursion of a complexity similar to the recursions adopted for maximum likelihood estimation within the EM algorithm, so that global decoding may be applied even for long sequences of data. 

\subsection{Basic latent Markov model}\label{sec:basicLM}
Following \cite{bart:farc:penn:13}, and as already mentioned above, the basic LM model rules out individual covariates and assumes that the conditional response probabilities are time homogenous.
In symbols, we have that $\phi_{y|u\bl x}^{(t)}=\phi_{y|u}$ in the univariate case and $\phi_{jy|u\bl x}^{(t)}=\phi_{jy|u}$ in the multivariate case; we also have $\pi_{u|\bl x}=\pi_u$ and $\pi_{u|\bu\bl x}^{(t)}=\pi_{u|\bu}^{(t)}$.

In order to fit this model, we can use function \code{est\_lm\_basic}, which is one of the main functions in the \pkg{LMest} package.
Here, we skip a detailed illustration of this function, as it is already provided in the Appendix of the book \cite{bart:farc:penn:13}, and we prefer to focus on functions to estimate more sophisticated LM versions which are only recently available in the package.
These versions include individual covariates and further latent variables to cluster subjects in different subpopulations (mixed LM model).
\section{Covariates in the measurement model}\label{sec:measurement_cov}
When the individual covariates are included in the measurement model, the conditional distribution of the response variables given the latent states 
may be parameterized by generalized logits.
In such a situation, the latent variables account for the unobserved heterogeneity, that is, the heterogeneity between subjects that we cannot explain on the basis of the observable covariates.
The advantage with respect to a standard random-effects or latent class
model with covariates is that the effect of the latent variables for the unobservable heterogeneity is allowed to be time-varying; for a further discussion see \cite{bart:farc:09} and \cite{penn:vitt:13}.
\subsection{Assumptions}\label{sec:measurement_cov:assumptions}
In dealing  with univariate data in which each response variable has an ordinal nature as in the following illustrative example, we denote the number of its response categories by $c$.
By retaining the assumption of local independence, and given the nature of the response variable, we rely on a parameterization based on global logits of the following type:
\begin{equation}\label{eq:cov2}
\log \frac{p(Y^{(t)}\geq y|U^{(t)}= u,\b X^{(t)}= \b x)}{p(Y^{(t)}< y|U^{(t)}=u ,\b X^{(t)}=\b x)} =
 \log\frac{\phi_{y|u\bl x}^{(t)}+\ldots+\phi_{c-1|u\bl x}^{(t)}}{\phi_{0|u\bl x}^{(t)}+\ldots +\phi_{y-1|u\bl x}^{(t)}}=\mu_y + \al_u + \b x\tr \b \beta,
\end{equation}
with $t=1,\ldots,T$, $y=1,\ldots,c-1$. 
Note that these logits reduce to standard logits in the case of binary variables, that is, when $c=2$.
In the above expression, $\mu_y$ are cut-points, $\al_u$ are the support points corresponding to each latent state, and $\b \beta$ is the vector of regression parameters for the covariates.

As mentioned in Section \ref{sec:LMgeneral}, the inclusion of individual covariates in the measurement model is typically combined with the constraints that $\pi_{u|\bl x}=\pi_u$ and $\pi_{u|\bu,\bl x}^{(t)}=\pi_{u|\bu}$, $t=1,\ldots,T$, $\bu,u = 1,\ldots,k$, in order to avoid interpretability problems of the resulting model.
Also note that, under these constraints, the transition probabilities are assumed to be time-homogeneous so as to reduce the number of free parameters. 
\subsection{Application to the health related longitudinal data} \label{ex_1}
The LM model with individual covariates in the measurement model, and then affecting the conditional distribution of the response variables given the latent process, is estimated by the function \code{est\_lm\_cov\_manifest} available in the 
\pkg{LMest} package.
In this section, we illustrate the use of this function on the basis of an application to data derived form the Health and Retirement Study (HRS) conducted by the University of Michigan\footnote{For more details on the study see \url{http://hrsonline.isr.umich.edu/}}. 
The data concern aspects related to retirement and health among elderly individuals in the USA.
The sampling design is nationally representative of the population over age 50 years, whereas the response variable is the self-reported health status (named SHLT) and it is measured on a scale based on five ordered categories: 
`poor', `fair', `good', `very good', `excellent'. The sample includes   
 $n = 7074 $ individuals followed at $ T = 8 $ approximately equally spaced occasions. 
Since each wave is added two years after the previous one, unobserved factors which vary during the course of the study may potentially affect the health status. 
The LM model with covariates directly takes this issue into account. 
  
For every subject, the available covariates are: gender, race, education, and age.  The  
data matrix consists of a number of rows equal to the number of subjects,  $n=7074$, and $19$ columns as follows: 

\begin{knitrout}
\definecolor{shadecolor}{rgb}{0.969, 0.969, 0.969}\color{fgcolor}\begin{kframe}
\begin{alltt}
\hlkwd{library}\hlstd{(}\hlstr{'LMest'}\hlstd{)}
\end{alltt}
\end{kframe}
\end{knitrout}

\begin{knitrout}
\definecolor{shadecolor}{rgb}{0.969, 0.969, 0.969}\color{fgcolor}\begin{kframe}
\begin{alltt}
\hlkwd{data}\hlstd{(data_SRHS)}
\hlstd{data} \hlkwb{=} \hlstd{data_SRHS}
\hlstd{data} \hlkwb{=} \hlkwd{as.matrix}\hlstd{(data)}
\hlkwd{head}\hlstd{(data)}
\end{alltt}
\begin{verbatim}
##      R1SHLT R2SHLT R3SHLT R4SHLT R5SHLT R6SHLT R7SHLT R8SHLT RAGENDER
## [1,]      4      4      3      3      4      3      3      3        1
## [2,]      3      3      3      3      3      3      3      3        2
## [3,]      1      1      2      3      3      2      3      4        1
## [4,]      3      3      4      4      3      1      3      5        2
## [5,]      1      1      1      1      1      1      1      2        2
## [6,]      3      2      2      2      2      2      2      2        2
##      RARACEM RAEDUC R1AGEY_E R2AGEY_E R3AGEY_E R4AGEY_E R5AGEY_E R6AGEY_E
## [1,]       1      3       56       58       60       62       64       66
## [2,]       1      5       54       55       57       59       61       63
## [3,]       1      3       53       55       57       58       60       62
## [4,]       1      5       36       38       40       42       44       46
## [5,]       1      3       46       48       50       52       54       56
## [6,]       1      4       44       47       48       50       52       54
##      R7AGEY_E R8AGEY_E
## [1,]       68       70
## [2,]       65       67
## [3,]       64       66
## [4,]       48       50
## [5,]       58       60
## [6,]       56       58
\end{verbatim}
\end{kframe}
\end{knitrout}
We observe that the first eight columns are related to the responses concerning the categorical response variable observed
at each time occasion, with $c=5$ and $r=1$. 
The ordered categories are originally coded from 1 to 5.  
The remaining columns are related to the covariates. 
 In particular, gender is coded as 1 for male and 2 for female,  whereas race has three categories coded as 1 for white, 2 for black, and 3 for others.
Finally, educational level is represented by five ordered categories coded as 1 for high school,  2  for general educational diploma,  3 for high school graduate,  4 for some college,  5 for college and above. The last eight columns are related to age
 recorded at each time occasion.

The function \code{est\_lm\_cov\_manifest} is used to estimate the LM model at issue; it requires the following main arguments as input:
\begin{itemize}
\item \code{S}: design array for the response configurations (of dimension $n\times T \times r$);
\item \code{X}: array of covariates; 
\item \code{lev}: number of response categories of each outcome; 
\item \code{k}: number of latent states; 
\item \code{mod}: type of model to be estimated, coded as \code{mod=0} for the model illustrated in Section \ref{sec:measurement_cov:assumptions} based on parameterization (\ref{eq:cov2}). In such a context, the latent process  is of first order with initial
probabilities which coincide with those of the stationary distribution of the chain. When  \code{mod=1} the function 
estimates a model relying on the assumption that the latent process has a distribution given by a mixture of AR(1) processes with common variance $\sigma^2$ and specific correlation coefficients $\rho_u$. 
This model strictly follows the one proposed by \cite{bart:bacc:penn:14}; see also \cite{penn:vitt:13} for a comparison between the two types of model and  the function help for further details. 
\item \code{q}: number of support points of the AR structure. When \code{mod=0} it has to be set equal to 1;
\item \code{tol}: tolerance level for the convergence, which corresponds to $\epsilon$ in   definition (\ref{eq:tol1}). The default value is \code{10e-8};
\item \code{maxit}: maximum number of iterations before convergence. The default value is \code{1000};
\item \code{start}: equal to \code{0}  for deterministic starting values (default value) and to \code{1} for random starting values;
\item \code{output}: equal to \code{TRUE} to print additional output. \code{FALSE} is the default option; 
\item  \code{out\_se}: equal to \code{TRUE} to calculate the information matrix and the standard errors. \code{FALSE} is the default option.
\end{itemize}
Function \code{est\_lm\_cov\_manifest} has other arguments for which we refer to the help. 

The matrix of the responses \code{S} is referred to $n = 7074$ subjects observed at $T = 8$ time occasions with respect to $r=1$ response variable. 
The responses are coded from 0 (``poor") to 4 (``excellent"), after modifying the original coding. 
For example, for the data at hand, the individual with  \code{id == 5} provides responses ``good" or ``excellent" at each time occasion:

\begin{knitrout}
\definecolor{shadecolor}{rgb}{0.969, 0.969, 0.969}\color{fgcolor}\begin{kframe}
\begin{alltt}
\hlstd{S} \hlkwb{=} \hlnum{5}\hlopt{-}\hlstd{data[,}\hlnum{1}\hlopt{:}\hlstd{TT]}
\hlkwd{head}\hlstd{(S)}
\end{alltt}
\begin{verbatim}
##      R1SHLT R2SHLT R3SHLT R4SHLT R5SHLT R6SHLT R7SHLT R8SHLT
## [1,]      1      1      2      2      1      2      2      2
## [2,]      2      2      2      2      2      2      2      2
## [3,]      4      4      3      2      2      3      2      1
## [4,]      2      2      1      1      2      4      2      0
## [5,]      4      4      4      4      4      4      4      3
## [6,]      2      3      3      3      3      3      3      3
\end{verbatim}
\end{kframe}
\end{knitrout}

In this application, we consider the covariates gender (included by a dummy variable equal to 1 for females),
race (by a dummy variable equal to 1 for non-white people), educational level (by two dummy variables: the first equal to 1 for some college education and the second equal to 1 for college education and above), age (scaled by 50), and age squared (again scaled by 50). 
For example, the covariates configuration of individual with  \code{id == 3994}, at occasion $t=1$, is:  

\begin{knitrout}
\definecolor{shadecolor}{rgb}{0.969, 0.969, 0.969}\color{fgcolor}\begin{kframe}
\begin{alltt}
\hlstd{X[,,}\hlnum{3994}\hlstd{,}\hlnum{1}\hlstd{]}
\end{alltt}
\begin{verbatim}
##      [,1] [,2] [,3] [,4] [,5] [,6]
## [1,]    1    0    0    0    1    1
## [2,]    1    0    0    0    1    1
## [3,]    1    0    0    0    1    1
## [4,]    1    0    0    0    1    1
\end{verbatim}
\end{kframe}
\end{knitrout}

The above subject is a female, white (first and second columns),
with high school diploma (third and fourth columns), begin 50 years old at the first interview (fifth column).  The last column is the square of age minus 50. 
Note that an alternative parameterization, which may result in a simpler interpretation of the parameter estimates, is formulated by considering as covariates the baseline age and the time since the beginning of the longitudinal study.

In order to obtain estimates for the data reported above, we use the following command in \proglang{R}: 

\begin{knitrout}
\definecolor{shadecolor}{rgb}{0.969, 0.969, 0.969}\color{fgcolor}\begin{kframe}
\begin{alltt}
\hlkwd{est_lm_cov_manifest}\hlstd{(S, X,} \hlkwc{lev} \hlstd{=} \hlnum{5}\hlstd{,} \hlkwc{k} \hlstd{=} \hlnum{10}\hlstd{,} \hlkwc{q} \hlstd{=} \hlnum{1}\hlstd{,} \hlkwc{mod} \hlstd{=} \hlnum{0}\hlstd{,} \hlkwc{tol} \hlstd{=} \hlnum{10e-08}\hlstd{,} 
                    \hlkwc{start} \hlstd{=} \hlnum{1}\hlstd{,} \hlkwc{output} \hlstd{=} \hlnum{TRUE}\hlstd{,} \hlkwc{out_se} \hlstd{=} \hlnum{TRUE}\hlstd{)}
\end{alltt}
\end{kframe}
\end{knitrout}

In such an application, we note that when $k$ is large the log-likelihood presents several local maxima. 
In order to prevent the problem of the multimodality of the log-likelihood we adopt both deterministic (by using \code{start=0}) and random (\code{start=1}) initializations of the EM algorithm. In particular, the random initialization
is performed on the basis of a suitable number of starting values proportional to $k$.
Then, we run the above code with a number of states $k$ from 1 to 11.
For each $k$, we consider the largest value of the log-likelihood at convergence and we compute the $BIC$ so as to choose the number of latent states.  
For the data at hand, the model which results to have the minimum $BIC$ is that with $k=10$ latent states. 

Note that the option  \code{out\_se = TRUE} is used for obtaining the standard errors for the parameter estimates, by means of the numerical method described in Section \ref{sec:EM}, so as to evaluate the effect of the covariates on  the probability of responding with a certain category. 

In the following we show the results obtained for the model with $k=10$ latent states, by using the \code{print} function, which also returns the main convergence info:

\begin{knitrout}
\definecolor{shadecolor}{rgb}{0.969, 0.969, 0.969}\color{fgcolor}\begin{kframe}
\begin{alltt}
\hlkwd{print}\hlstd{(final_new10)}
\end	{alltt}
\begin{verbatim}
## Call:
## est_lm_cov_manifest(S = S, X = X, lev = 5, k = 10, q = 1, mod = 0, 
##     tol = 1e-08, start = 1, output = TRUE, out_se = TRUE)
## 
## Convergence info:
##      LogLik  np    AIC    BIC
## [1,] -62579 109 125376 126124
\end{verbatim}
\end{kframe}
\end{knitrout}

We also report the results of  the \code{summary} output for the chosen LM model: 

\begin{knitrout}
\definecolor{shadecolor}{rgb}{0.969, 0.969, 0.969}\color{fgcolor}\begin{kframe}
\begin{alltt}
\hlkwd{summary}\hlstd{(final_new10)}
\end{alltt}
\begin{verbatim}
## Call:
## est_lm_cov_manifest(S = S, X = X, lev = 5, k = 10, q = 1, mod = 0, 
##     tol = 1e-08, start = 1, output = TRUE, out_se = TRUE)
## 
## Coefficients:
## 
##   Vector of cutpoints:
## [1]  8.284  4.543  0.747 -3.573
## 
##  Support points for the latent states:
##  [1] -12.179  -7.279  -4.472  -2.279  -1.179  -0.158   2.176   4.451
##  [9]   5.309   7.532
## 
##  Estimate of the vector of regression parameters:
## [1] -0.226 -1.424  1.534  2.643 -0.124  0.000
## 
##  Vector of initial probabilities:
##  [1] 0.2219 0.0932 0.0465 0.0644 0.0256 0.0094 0.2180 0.0220 0.2849 0.0142
## 
##  Transition matrix:
##         [,1]   [,2]   [,3]   [,4]   [,5]   [,6]   [,7]   [,8]   [,9]
##  [1,] 0.8965 0.0040 0.0111 0.0257 0.0000 0.0019 0.0083 0.0056 0.0384
##  [2,] 0.0006 0.8356 0.0000 0.0033 0.0299 0.0146 0.1150 0.0000 0.0010
##  [3,] 0.0371 0.0000 0.7998 0.0641 0.0000 0.0000 0.0003 0.0745 0.0103
##  [4,] 0.0439 0.0056 0.0618 0.7700 0.0000 0.0031 0.0261 0.0393 0.0272
##  [5,] 0.0001 0.1474 0.0000 0.0002 0.8308 0.0171 0.0042 0.0000 0.0001
##  [6,] 0.0250 0.1922 0.0001 0.0507 0.0792 0.5893 0.0582 0.0000 0.0050
##  [7,] 0.0011 0.0244 0.0000 0.0024 0.0034 0.0047 0.8691 0.0000 0.0688
##  [8,] 0.1073 0.0000 0.1046 0.0998 0.0000 0.0000 0.0001 0.6591 0.0184
##  [9,] 0.0483 0.0099 0.0004 0.0052 0.0002 0.0014 0.0358 0.0004 0.8911
## [10,] 0.1247 0.0266 0.0317 0.0789 0.0000 0.0000 0.2421 0.0098 0.3337
##        [,10]
##  [1,] 0.0084
##  [2,] 0.0000
##  [3,] 0.0138
##  [4,] 0.0229
##  [5,] 0.0000
##  [6,] 0.0004
##  [7,] 0.0262
##  [8,] 0.0105
##  [9,] 0.0074
## [10,] 0.1524
## 
##  Standard errors for the regression parameters:
## [1] 0.0762 0.1039 0.1061 0.0925 0.0068 0.0002
\end{verbatim}
\end{kframe}
\end{knitrout}

The above output displays the estimated cut-points $\hat{\mu}_y$, the estimated support points $\hat{\alpha}_u$, and the estimated vector of regression parameters  $\hat{\b \beta}$ as in expression (\ref{eq:cov2}). 
The estimated coefficients in $\hat{\b \beta}$ are reported in the same order adopted to define the matrix \code{X} of covariates.
The list of objects returned by the function, contained in
\code{final\_new10}, may also be displayed in the usual way; for a complete
list of the arguments returned as output we refer to the package help.  
For example, the standard errors for the  estimated  regression coefficients 
may be obtained as

\begin{knitrout}
\definecolor{shadecolor}{rgb}{0.969, 0.969, 0.969}\color{fgcolor}\begin{kframe}
\begin{alltt}
\hlkwd{round}\hlstd{(final_new10}\hlopt{$}\hlstd{sebe,}\hlnum{3}\hlstd{)}
\end{alltt}
\begin{verbatim}
## [1] 0.076 0.104 0.106 0.093 0.007 0.000
\end{verbatim}
\end{kframe}
\end{knitrout}

On the basis of the above results, we can evaluate the effect of the covariates on  the probability of reporting a certain level of the health 
status. In particular, women tend to report worse health status than men (the odds ratio for females versus males is equal to $\exp(-0.226) = 0.798$), whereas white individuals have a higher probability of reporting a good health status with respect to non-white ones (the odds ratio for non-white versus white is equal to $\exp(-1.424) = 0.241$).
We also observe that better educated individuals tend to have a higher opinion about their health status.
Finally, the effect of age is negative and its trend is linear due to the fact that the quadratic term of age is not significant.  

In this example, the time-varying random effects are used to account for the unobserved heterogeneity and interpretation of the latent distribution is not of primary interest.
The fact that the optimal number of states is $k=10$ tells us that there is unobserved heterogeneity, that is, that self-reported health status cannot be completely explained merely by the few covariates we have used.
For another application of the LM model to the ordinal longitudinal data see \cite{penn:vitt:13}. 
\section{Covariates in the latent model}
\label{sec:covlat}

When the covariates are included in the latent model, we suppose that the response variables 
measure and depend on an individual characteristic of interest (e.g. the quality of life), 
which may evolve over time and is not directly observable. 
In such a case, the main research interest is in modeling the effect of covariates 
on the latent trait distribution \citep{bart:lupp:mont:09}.
This is illustrated in the following.

\subsection{Assumptions}\label{sec:modLat}
A natural way to allow the initial and transition probabilities of the latent Markov chain to depend on individual covariates is by adopting a multinomial logit parameterization as follows:
\begin{eqnarray}  \label{eq:be}
\log \frac{p(U^{(1)}=u|\b X^{(1)}=\b x)}{p(U^{(1)}=1|\b X^{(1)}=\b x)} &=& \log \frac{\pi_{u|\bl x}}{\pi_{1|\bl x}}=
\beta_{0u}+\b x\tr \b \beta_{1u},\quad u\geq 2,\\ \label{eq:Ga}
\log \frac{p(U^{(t)}=u |U^{(t-1)}=\bu,\b X^{(t)}=\b x)}
{p(U^{(t)}=\bu|U^{(t-1)}=\bu,\b X^{(t)}=\b x)}&=&\log \frac{\pi_{u|\bu\bl x}^{(t)}}{\pi_{\bu|\bu\bl x}^{(t)}} = \gamma_{0\bu u}+\b x\tr
\b\gamma_{1\bu u}, \quad t\geq 2,\;\;\bu\neq u. 
\end{eqnarray}
In the above expressions, $\b\beta_u=(\beta_{0u},\b\beta_{1u}\tr)\tr$ and $\b\gamma_{uv}=(\gamma_{0\bu u},\b\gamma_{1\bu u}\tr)\tr$ are parameter vectors to be estimated which are collected in the matrices $\b\beta$ and  $\b\Gamma$.

For a  more parsimonious model, instead of using (\ref{eq:Ga}) we can rely on the following parameterization for the transition probabilities, that is, a multinomial logit based on the difference between two sets of parameters:
\begin{equation}\label{eq:Gadiff}
\log \frac{p(U^{(t)}=u|U^{(t-1)}=\bu,\b X^{(t)}=\b x)}
{p(U^{(t)}=\bu|U^{(t-1)}=\bu,\b X^{(t)}=\b x)} =\log \frac{\pi_{u|\bu\bl x}^{(t)}}{\pi_{\bu|\bu\bl x}^{(t)}} =
\gamma_{0uv}+\b x\tr (\b\gamma_{1\bu}-\b\gamma_{1u}),
\end{equation}
with $t \geq 2$ and $\bu\neq u$ and where $\b\gamma_{11}=0$ to ensure model identifiability.
The parameterization used for modeling the initial probabilities is again based on standard multinomial logits, as defined in (\ref{eq:be}).

As already mentioned, when the covariates affect the distribution of the latent process, these covariates are typically excluded from the measurement model by means of the constraint $\phi_{y|u\bl x}^{(t)}=\phi_{y|u}$ in the univariate case or $\phi_{jy|u\bl x}^{(t)}=\phi_{jy|u}$ in the multivariate case.
In this case, we rely on the assumption of time homogeneity for the conditional response probabilities which are also independent of the covariates.

Both parameterizations above, based on (\ref{eq:Ga}) and (\ref{eq:Gadiff}), are implemented in the \proglang{R} function  \code{est\_lm\_cov\_latent}, which allows us to estimate the resulting LM models as we describe in detail below.
\subsection{Application to the health related longitudinal data}
To illustrate function \code{est\_lm\_cov\_latent} we consider the same data as in Section \ref{ex_1}.
In such a context, an interesting research question is about the relationship between the self-reported health status and the covariates.
When the latter are included in the latent model, the initial and transition probabilities are estimated for each configuration of the covariates and this may be useful to identify clusters of subjects related to specific needs. 

The \proglang{R} function is based on the following main arguments as input: 
\begin{itemize}
\item \code{S}: array of available configurations (of dimension
  $n\times T \times r$); 
\item \code{X1}: matrix of covariates affecting the initial probabilities; 
 \item \code{X2}: array of covariates affecting the transition probabilities;
\item \code{yv}: vector of frequencies of the available covariate and response configurations;
\item \code{k}: number of latent states; 
\item \code{start}: equal to \code{0} for deterministic starting values (default), to \code{1} for random starting values, and to \code{2} to define initial values as input of the function; 
\item \code{tol}: tolerance level for the convergence. The default value is \code{10e-8};
\item \code{maxit}: maximum number of iterations before convergence. The default value is \code{1000};
\item \code{param}: type of parameterization for the transition probabilities, coded as \code{param = "multilogit"} (default) to allow for the model parameterization defined in  (\ref{eq:Ga}) and as  \code{param = "difflogit"} to allow for the parameterization defined in (\ref{eq:Gadiff}); 
\item \code{Psi, Be, Ga}: 
initial values of the matrix of the conditional response probabilities and of the parameters affecting the logits for the initial and transition probabilities when \code{start = 2};
\item  \code{output}: equal to \code{TRUE} to print additional output. \code{FALSE} is the default option; 
\item  \code{out\_se}:  equal to  \code{TRUE}  to compute the information matrix and standard errors. \code{FALSE} is the default option; 
\item \code{fixPsi}: equal to \code{TRUE} if the matrix of conditional response probabilities is given in input and it is not updated anymore. \code{FALSE} is the default option. 
\end{itemize}

The data matrix is specified as in Section \ref{ex_1} and the \proglang{R} function requires the array \code{S} of available response configurations, again corresponding to $n=7074$ subjects, $T=8$ time occasions, and with $r=1$ response variable. In this case, the vector of frequencies \code{yv} is an all-ones vector of length $n$. As an example, the individual with \code{id==3994} provides the following responses:

\begin{knitrout}
\definecolor{shadecolor}{rgb}{0.969, 0.969, 0.969}\color{fgcolor}\begin{kframe}
\begin{alltt}
\hlstd{S[}\hlnum{3994}\hlstd{,]}
\end{alltt}
\begin{verbatim}
## R1SHLT R2SHLT R3SHLT R4SHLT R5SHLT R6SHLT R7SHLT R8SHLT 
##      3      3      4      3      3      3      3      4
\end{verbatim}
\end{kframe}
\end{knitrout}

The model is specified through the design matrices for the initial and transition probabilities. 
More in detail, it is necessary to generate the design matrix \code{X1} containing the covariates affecting the initial probabilities of the latent process. 
In the present application, we have $n = 7074$ subjects and 6 covariates, defined for the first time occasion as:

\begin{knitrout}
\definecolor{shadecolor}{rgb}{0.969, 0.969, 0.969}\color{fgcolor}\begin{kframe}
\begin{alltt}
 \hlkwd{head}\hlstd{(X1)}
\end{alltt}
\begin{verbatim}
##      [,1] [,2] [,3] [,4] [,5] [,6]
## [1,]    0    0    0    0    6 0.36
## [2,]    1    0    0    1    4 0.16
## [3,]    0    0    0    0    3 0.09
## [4,]    1    0    0    1  -14 1.96
## [5,]    1    0    0    0   -4 0.16
## [6,]    1    0    1    0   -6 0.36
\end{verbatim}
\end{kframe}
\end{knitrout}

Note that the covariates are coded as in the previous example in Section \ref{ex_1} (the square of age is divided by 100).
Accordingly, for $t=2,\ldots,T$, the design matrix for the covariates affecting the transition probabilities of the latent process, \code{X2},  is generated by
considering $n  = 7074$ subjects, $T-1 = 7$ time occasions, and 6 covariates. 
For example, for the same individual as above with \code{id==3994} the matrix is defined as

\begin{knitrout}
\definecolor{shadecolor}{rgb}{0.969, 0.969, 0.969}\color{fgcolor}\begin{kframe}
\begin{alltt}
 \hlstd{X2[}\hlnum{3994}\hlstd{,,]}
\end{alltt}
\begin{verbatim}
##      [,1] [,2] [,3] [,4] [,5] [,6]
## [1,]    1    0    0    0    3 0.09
## [2,]    1    0    0    0    5 0.25
## [3,]    1    0    0    0    7 0.49
## [4,]    1    0    0    0    9 0.81
## [5,]    1    0    0    0   11 1.21
## [6,]    1    0    0    0   13 1.69
## [7,]    1    0    0    0   15 2.25
\end{verbatim}
\end{kframe}
\end{knitrout}

From the matrix above, we observe that the subject is a white female with high school diploma, who was 53 years old at the second interview.

Within this illustrative example, we fit the model defined in Section \ref{sec:modLat}, with a number of latent states equal to the number of response categories, that is, $k=5$, by using the following command in \proglang{R}:

\begin{knitrout}
\definecolor{shadecolor}{rgb}{0.969, 0.969, 0.969}\color{fgcolor}\begin{kframe}
\begin{alltt}
\hlstd{res_5} \hlkwb{=} \hlkwd{est_lm_cov_latent}\hlstd{(}\hlkwc{S} \hlstd{= S,} \hlkwc{X1} \hlstd{= X1,} \hlkwc{X2} \hlstd{= X2,} \hlkwc{k} \hlstd{=} \hlnum{5}\hlstd{,} \hlkwc{start} \hlstd{=} \hlnum{0}\hlstd{,}
               \hlkwc{param} \hlstd{=} \hlstr{"multilogit"}\hlstd{,} \hlkwc{output} \hlstd{=} \hlnum{TRUE}\hlstd{)}
\end{alltt}
\end{kframe}
\end{knitrout}

The results can be displayed using the functions \code{print} and
\code{summary}. 
The  \code{print} function provides information about log-likelihood, AIC, and BIC:

\begin{knitrout}
\definecolor{shadecolor}{rgb}{0.969, 0.969, 0.969}\color{fgcolor}\begin{kframe}
\begin{alltt}
\hlkwd{print}\hlstd{(res_5)}
\end{alltt}
\begin{verbatim}
## Call:
## est_lm_cov_latent(S = S, X1 = X1, X2 = X2, k = 5, start = 0, 
##     param = "multilogit", fort = TRUE, output = TRUE)
## 
## Convergence info:
##      LogLik  np    AIC    BIC
## [1,] -62427 188 125229 126520
\end{verbatim}
\end{kframe}
\end{knitrout}

The \code{summary} function displays the main objects: 

\begin{knitrout}
\definecolor{shadecolor}{rgb}{0.969, 0.969, 0.969}\color{fgcolor}\begin{kframe}
\begin{alltt}
\hlkwd{summary}\hlstd{(res_5)}
\end{alltt}
\begin{verbatim}
## Call:
## est_lm_cov_latent(S = S, X1 = X1, X2 = X2, k = 5, start = 0, 
##     param = "multilogit", fort = TRUE, output = TRUE)
## 
## Coefficients:
## 
##  Be - Parameters affecting the logit for the initial probabilities:
##            logit
##                     2        3        4        5
##   intercept  0.736261  1.71894  1.60361  1.61932
##   X11       -0.004668 -0.29976 -0.10427 -0.23419
##   X12        0.020255 -0.37203 -1.14359 -1.43806
##   X13        0.449647  1.12496  1.49687  1.76936
##   X14        0.291721  1.70277  2.48762  3.00167
##   X15       -0.035662 -0.03377 -0.04263 -0.07273
##   X16        0.316996  0.21178  0.20712  0.24153
## 
##  Ga - Parameters affecting the logit for the transition probabilities:
## , , logit = 1
## 
##            logit
##                   2        3        4        5
##   intercept -3.4506 -30.3104 -7.32596  -0.7017
##   X21       -0.1625 -10.1897 -1.01965  -6.8384
##   X22       -0.2798  -8.4566  1.11113 -10.7134
##   X23       -0.1829   0.3931 -8.20077  -1.3429
##   X24        0.7692   2.3875 -7.43413  -9.0080
##   X25        0.3068   1.8111  0.13874  -0.4990
##   X26       -1.4401  -2.9503 -0.03292   0.9238
## 
## , , logit = 2
## 
##            logit
##                    2        3        4         5
##   intercept -3.06862 -2.17493 -16.3128 -14.54901
##   X21        0.25111 -0.17557  -1.4647  -0.44729
##   X22       -0.30635 -0.69504   0.5331   9.38099
##   X23        0.04224  0.54140  -8.4763  -6.88290
##   X24        0.28653 -0.01384  -2.5061  -6.96501
##   X25        0.02297 -0.04573   1.7420   0.02022
##   X26       -0.06965  0.06330  -5.9457   0.09913
## 
## , , logit = 3
## 
##            logit
##                    2        3         4        5
##   intercept -4.61886 -2.00810 -3.438446 -4.03956
##   X21       -0.45067 -0.26236 -0.190092 -1.89834
##   X22        0.26960 -0.06015 -0.003967  2.71185
##   X23       -1.79197 -0.22210 -0.635222 -0.58401
##   X24       -0.50820 -0.62816 -1.030221 -1.71889
##   X25       -0.03976 -0.03865  0.022347 -0.08331
##   X26        0.32190  0.16203  0.007955 -0.18049
## 
## , , logit = 4
## 
##            logit
##                    2        3         4         5
##   intercept -6.99631 -6.00784 -2.166489 -2.965711
##   X21       -0.55850  0.21417 -0.165402 -0.756391
##   X22        0.85805  0.92151  0.390139  0.007904
##   X23        0.85650 -0.48585 -0.206565 -0.396467
##   X24       -1.09054  0.02140 -0.443345 -1.582536
##   X25        0.06578 -0.04761  0.007509 -0.088671
##   X26        0.13126  0.61353  0.028437  0.311526
## 
## , , logit = 5
## 
##            logit
##                   2        3        4        5
##   intercept -16.217 -2.24574 -2.21537 -1.35550
##   X21         1.429 -2.30980 -0.75273 -0.20117
##   X22         3.383  0.72911  2.04474  0.19405
##   X23        -6.884 -1.98413 -1.56324 -0.07730
##   X24        -6.674 -4.19165 -1.69996 -0.38578
##   X25         1.678  0.03181 -0.03745 -0.01392
##   X26        -8.260 -1.19747  0.03455  0.05452
## 
## 
##  Psi - Conditional response probabilities:
## , , item = 1
## 
##         state
## category        1        2        3        4         5
##        0 0.698132 0.059749 0.004321 0.001702 0.0004021
##        1 0.266964 0.682778 0.084389 0.008304 0.0013713
##        2 0.025484 0.220297 0.713140 0.145421 0.0313998
##        3 0.008256 0.028767 0.181283 0.756656 0.1871161
##        4 0.001165 0.008409 0.016866 0.087918 0.7797108
\end{verbatim}
\end{kframe}
\end{knitrout}

On the basis of the estimated conditional response probabilities displayed above, we observe that the latent states are ordered: subjects classified in the first latent state have higher probability of reporting the worst health status, whereas subjects classified in the last state have a higher probability of reporting the best health status. 

The parameter matrix \code{Be} has large and positive intercepts,
indicating a general tendency to start in good health status. Gender
log-odds (second row of \code{Be}) are all negative, indicating that females report
worse health status than males at the first occasion. The fourth and
five rows give large positive parameters, indicating that better
education leads to better health. Finally, the negative parameters for
age (sixth row) indicate that elder subjects generally start into
a worse health status. 

By using the option \code{output=TRUE} the function also returns some additional outputs.
In particular, it is possible to obtain the estimated initial probability matrix, \code{Piv}, and the estimated transition probabilities matrix,
\code{PI}. Accordingly, it may be of interest to compute the average initial and transition probability matrices for a group of subjects of 
interest. For example, if we consider white females with high school diploma, we obtain the corresponding
average estimated initial and transition probabilities by means of the following commands:

\begin{knitrout}
\definecolor{shadecolor}{rgb}{0.969, 0.969, 0.969}\color{fgcolor}\begin{kframe}
\begin{alltt}
\hlstd{ind1} \hlkwb{=} \hlstd{(data[,}\hlnum{9}\hlstd{]}\hlopt{==}\hlnum{2} \hlopt{&} \hlstd{data[,}\hlnum{10}\hlstd{]}\hlopt{==}\hlnum{1} \hlopt{&} \hlstd{data[,}\hlnum{11}\hlstd{]}\hlopt{==}\hlnum{1}\hlstd{)}
\hlstd{piv1} \hlkwb{=} \hlkwd{round}\hlstd{(}\hlkwd{colMeans}\hlstd{(res_5}\hlopt{$}\hlstd{Piv[ind1,]),}\hlnum{4}\hlstd{)}
\hlkwd{print}\hlstd{(piv1)}
\end{alltt}
\begin{verbatim}
##      1      2      3      4      5 
## 0.0705 0.1438 0.2736 0.2841 0.2280
\end{verbatim}
\end{kframe}
\end{knitrout}

\begin{knitrout}
\definecolor{shadecolor}{rgb}{0.969, 0.969, 0.969}\color{fgcolor}\begin{kframe}
\begin{alltt}
\hlstd{PI1} \hlkwb{=} \hlkwd{round}\hlstd{(}\hlkwd{apply}\hlstd{(res_5}\hlopt{$}\hlstd{PI[,,ind1,}\hlnum{2}\hlopt{:}\hlstd{TT],}\hlkwd{c}\hlstd{(}\hlnum{1}\hlstd{,}\hlnum{2}\hlstd{),mean),}\hlnum{4}\hlstd{)}
\hlkwd{print}\hlstd{(PI1)}
\end{alltt}
\begin{verbatim}
##      state
## state      1      2      3      4      5
##     1 0.9150 0.0819 0.0000 0.0016 0.0015
##     2 0.0611 0.8811 0.0556 0.0023 0.0000
##     3 0.0067 0.0781 0.8827 0.0316 0.0008
##     4 0.0016 0.0076 0.0989 0.8782 0.0136
##     5 0.0005 0.0029 0.0286 0.1598 0.8082
\end{verbatim}
\end{kframe}
\end{knitrout}

In a similar way, it is possible to compute the average initial and transition probabilities for non-white females with high educational level:

\begin{knitrout}
\definecolor{shadecolor}{rgb}{0.969, 0.969, 0.969}\color{fgcolor}\begin{kframe}
\begin{alltt}
\hlstd{ind2} \hlkwb{=} \hlstd{(data[,}\hlnum{9}\hlstd{]}\hlopt{==}\hlnum{2} \hlopt{&} \hlstd{data[,}\hlnum{10}\hlstd{]}\hlopt{>=}\hlnum{2} \hlopt{&} \hlstd{data[,}\hlnum{11}\hlstd{]}\hlopt{==}\hlnum{1}\hlstd{)}
\hlstd{piv2} \hlkwb{=} \hlkwd{round}\hlstd{(}\hlkwd{colMeans}\hlstd{(res_5}\hlopt{$}\hlstd{Piv[ind2,]),}\hlnum{4}\hlstd{)}
\hlkwd{print}\hlstd{(piv2)}
\end{alltt}
\begin{verbatim}
##      1      2      3      4      5 
## 0.1286 0.2652 0.3424 0.1647 0.0990
\end{verbatim}
\end{kframe}
\end{knitrout}

\begin{knitrout}
\definecolor{shadecolor}{rgb}{0.969, 0.969, 0.969}\color{fgcolor}\begin{kframe}
\begin{alltt}
\hlstd{PI2} \hlkwb{=} \hlkwd{round}\hlstd{(}\hlkwd{apply}\hlstd{(res_5}\hlopt{$}\hlstd{PI[,,ind2,}\hlnum{2}\hlopt{:}\hlstd{TT],}\hlkwd{c}\hlstd{(}\hlnum{1}\hlstd{,}\hlnum{2}\hlstd{),mean),}\hlnum{4}\hlstd{)}
\hlkwd{print}\hlstd{(PI2)}
\end{alltt}
\begin{verbatim}
##      state
## state      1      2      3      4      5
##     1 0.9311 0.0641 0.0000 0.0047 0.0000
##     2 0.0468 0.9148 0.0289 0.0042 0.0053
##     3 0.0085 0.0727 0.8755 0.0313 0.0119
##     4 0.0035 0.0146 0.1381 0.8306 0.0132
##     5 0.0121 0.0045 0.1768 0.1559 0.6507
\end{verbatim}
\end{kframe}
\end{knitrout}

From the above results we observe that, at the beginning of the period of observation,
white females have a probability  of around 0.50 of being in the last two states corresponding
to very good or excellent health conditions, whereas non-white females have a higher 
probability of being in the first three states, corresponding to worse health conditions.
Moreover, the estimated transition probability matrices show a quite high persistence in the same latent state for both groups of individuals.
However, the lower diagonal elements of the transition matrix related to non-white females are higher than those related to white females, except for some transitions, showing a worse perception of their health status.
\subsubsection{Decoding}

Local and global decoding are implemented in the \proglang{R} function \code{decoding}, which allows us to predict the sequence of latent states for a certain sample unit on the basis of the output of the main functions for estimation in the package.
Function \code{decoding} requires certain arguments in input:
\begin{itemize}
\item \code{est}: output from the main functions for estimation (i.e., \code{est\_lm\_basic}, \code{est\_lm\_cov\_latent}, \code{est\_lm\_cov\_manifest}, or \code{est\_lm\_mixed} illustrated in the following);
\item \code{Y}: vector or matrix of responses;
 \item \code{X1 }: matrix of covariates affecting the initial probabilities (for \code{est\_lm\_cov\_latent}) or affecting the distribution of the responses (for \code{est\_lm\_cov\_manifest});
  \item\code{X2 }: array of covariates affecting the transition probabilities (for \code{est\_lm\_cov\_latent}).
\end{itemize}

For the application above, the most likely sequence of latent states for all sample units may be
obtained by the following command:

\begin{knitrout}
\definecolor{shadecolor}{rgb}{0.969, 0.969, 0.969}\color{fgcolor}\begin{kframe}
\begin{alltt}
\hlstd{out_5} \hlkwb{=} \hlkwd{decoding}\hlstd{(res_5,S,X1,X2)}
\hlkwd{head}\hlstd{(out_5}\hlopt{$}\hlstd{Ug)}
\end{alltt}
\begin{verbatim}
##      [,1] [,2] [,3] [,4] [,5] [,6] [,7] [,8]
## [1,]    2    2    3    3    3    3    3    3
## [2,]    3    3    3    3    3    3    3    3
## [3,]    5    5    4    3    3    3    3    3
## [4,]    3    3    2    2    2    2    2    2
## [5,]    5    5    5    5    5    5    5    5
## [6,]    4    4    4    4    4    4    4    4
\end{verbatim}
\end{kframe}
\end{knitrout}
For instance, we conclude that for the first subject there is only a transition, at the third time occasion, from the second to the 
third latent state.

\section{Mixed latent Markov model}\label{sec_mixed}
Another relevant extension of LM models may be formulated to take into account additional sources of (time-fixed) dependence in the data.
In this paper, we provide an illustration of the mixed LM model \citep{vand:lang:90} in which the parameters of the latent process are allowed to vary in different latent subpopulations defined by an additional discrete latent variable.
\subsection{Assumptions}
Let $U$ be a discrete latent variable, which is time invariant. 
The latent process is now denoted by $\b V = (V^{(1)},\ldots, V^{(T)})$ 
which substitutes the symbol $\b U$ used in the previous sections.  In such a context, the variables
in $\b V$ follow a first order Markov chain only conditionally on
$U$. 
This additional latent variable has $k_1$ support points (corresponding to latent classes) 
and mass probabilities denoted by $\lambda_{u}$, $u=1,\ldots,k_1$. 
Accordingly, we denote by $k_2$ the number of latent states, corresponding to the number
of support points of every latent variable $V^{(t)}$, $t=1,\ldots,T$. 

Note that, under this approach, which may be useful from a perspective
of clustering, the initial and transition probabilities of the latent Markov chain 
differ between sample units in a way that does not depend on the observable covariates. 

In such a context, parameters to be estimated are the conditional response probabilities,
which may be formulated as 
\[ \phi_{jy|v}=p(Y_j^{(t)}=y|V^{(t)}=v),  \quad t=1,\ldots,T,\; y=0,\ldots,c_j-1,\; v = 1,\ldots,k_2, 
\]
the initial probabilities
\[
\pi_{v|u} = p(V^{(1)} = v | U = u), \quad u =1,\ldots,k_1, \; v = 1,\ldots,k_2,
\]
and the transition probabilities
\[
\pi_{v|u\bar{v}} = p(V^{(t)}=v | U = u, V^{(t-1)}=\bar{v}),\quad u=1,\ldots,k_1, \; \bar{v},v = 1,\ldots,k_2.
\]

The model here illustrated relies on the assumption that the conditional
response probabilities and the transition probabilities are time-homogeneous.
Obviously, this formulation may be extended by also including observable covariates as illustrated in the previous sections; 
see \cite{bart:farc:penn:13} for a description.

The manifest distribution of $\tilde{\b Y}$ may be obtained by extending the rules
given in Section \ref{sec:basicLM}. 
In particular, the conditional distribution of $\b V$ given $U$ may be formulated as 
\[
p(\b V = \b v | U = u) = \pi_{v^{(1)}|u} \prod_{t>1}\pi_{v^{(t)}|u\bar{v}^{(t-1)}}, 
\]
where $\b v = (v^{(1)},\ldots,v^{(T)})$ denotes a realization of $\b V$.
Given the assumption of local independence that is maintained under this model, the conditional distribution of 
$\tilde{\b Y}$ given $U$ and $\b V$ reduces to
\[
p(\tilde{\b Y}=\tilde{\b y}| U=u, \b V = \b v) = p(\tilde{\b Y}=\tilde{\b y}| \b V = \b v)  = \prod_t \phi_{\bl y^{(t)}|v^{(t)}} = \prod_t \prod_j \phi_{jy_j^{(t)}|v^{(t)}},
\]
whereas the conditional distribution of $\tilde{\b Y}$ given $U$ is expressed as
\[
p(\tilde{\b Y}=\tilde{\b y} |U = u) = \sum_{\bl v} \pi_{v^{(1)}|u}\pi_{v^{(2)}|u v^{(1)}}\ldots \pi_{v^{(T)}|u v^{(T-1)}}\phi_{\bl y^{(1)}|v^{(1)}}\ldots\phi_{\bl y^{(T)}|v^{(T)}}.  
\]
Finally, the manifest distribution of $\tilde{\b Y}$ is now obtained by the following sum
\[
p(\tilde{\b y}) = p(\tilde{\b Y} = \tilde{\b y}) = \sum_{u=1}^{k_1}p(\tilde{\b Y}= \tilde{\b y} |U = u)\lambda_{u},
\]
which depends on the mass probabilities for the distribution of the latent variable $U$. 
Even in this case $p(\tilde{\b y})$ may be computed through a forward recursion \citep{baum:et:al:70}.

Referred to the maximum likelihood estimation of the mixed LM model formulated
above, we can extend the procedure illustrated in Section \ref{sec:EM}, where the complete 
data log-likelihood has the following expression:
\begin{eqnarray*}
\hspace*{-1.5cm}\ell^*(\b\th) &=& \sum_{j=1}^r \sum_{t=1}^T\sum_{v=1}^{k_2}\sum_{y=0}^{c_j-1} a_{jvy}^{(t)}\log \phi_{jy|v}
\nonumber\\
&& +\sum_{u=1}^{k_1}\Bigg[\sum_{v=1}^{k_2}b_{iuv}^{(1)}\log \pi_{v|u}+
\sum_{t=2}^T\sum_{\bar{v}=1}^{k_2}\sum_{v=1}^{k_2} b_{ u\bar{v}v}^{(t)}\log \pi_{v|u\bar{v}}\Bigg] +\sum_{u=1}^{k_1} c_u\log \lambda_{u}.\label{eq:comp_lk_cov_mixed}   
\end{eqnarray*}

In the previous expression, $a_{jvy}^{(t)}$ is the number of sample units that are in latent state $v$ at occasion $t$ and provide response $y$ to variable $j$.
Moreover, with reference to latent class $u$ and occasion $t$, $b_{uv}^{(t)}$ is number of sample units in latent state $v$, and  $b_{iu\bar{v}v}^{(t)}$ is the number of transitions from state $\bar{v}$ to state $v$.
Finally, $c_{u}$ is the overall number of sample units that are in latent class $u$.

\subsection{Application example on longitudinal data from criminology}
The mixed LM model is illustrated by using a simulated dataset (for reasons of privacy) which strictly follows the one
analyzed in  \cite{bart:penn:fran:07}; see also  \cite{fran:liu:Soot:10}, \cite{bart:farc:penn:13}, and \cite{penn:14}.
The data concern the complete conviction histories of
a cohort of offenders followed from the age of criminal responsibility,
10 years. The offense code has been reduced to 73 major offenses and they have been grouped according to the Research Development and Statistics Directorate (1998)\footnote{\url{http://discover.ukdataservice.ac.uk/catalogue/?sn=3935}} on the basis of the following main typologies: `violence against the person', `sexual offenses', `burglary, robbery', `theft and handling stolen goods', `fraud and forgery', `criminal damage', `drug offenses', `motoring offenses', and `other offenses'.

For the simulated data, we consider $n=10000$ subjects (including  
the proportion of non-offenders): 4800 females and 5200 males.     
We also consider $T = 6$ age bands with a five years window (10-15, 16-20, 21-25, 26-30, 31-35, and 36-40 years) and $r=10$ binary response variables. For each age band, every response variable is equal to 1 if the subject has been convicted for a crime of the corresponding offense group and to 0 otherwise.

Then, the data matrix, reported below in long format, has been simulated on the basis of the same parameter estimates
obtained in  \cite{bart:penn:fran:07}.

\begin{knitrout}
\definecolor{shadecolor}{rgb}{0.969, 0.969, 0.969}\color{fgcolor}\begin{kframe}
\begin{alltt}
\hlkwd{data}\hlstd{(data_criminal_sim)}
\hlstd{data} \hlkwb{=} \hlkwd{as.matrix}\hlstd{(data_criminal_sim)}
\hlkwd{head}\hlstd{(data)}
\end{alltt}
\begin{verbatim}
##      id sex time y1 y2 y3 y4 y5 y6 y7 y8 y9 y10
## [1,]  1   1    1  0  0  0  0  0  0  0  0  0   0
## [2,]  1   1    2  0  0  0  0  0  0  0  0  0   0
## [3,]  1   1    3  0  0  0  0  0  0  0  0  0   0
## [4,]  1   1    4  0  0  0  0  0  0  0  0  0   0
## [5,]  1   1    5  0  0  0  0  0  0  0  0  0   0
## [6,]  1   1    6  0  0  0  0  0  0  0  0  0   0
\end{verbatim}
\end{kframe}
\end{knitrout}

The first column of the data matrix is the id code of the subject, whereas the covariate gender (second column) is coded
as 1 for male and 2 for female, 
the column named \code{time}  is referred to  the age bands, and the last ten columns are related to the binary response variables. 
In order to prepare the data, we run function \code{long2wide}, included in the package,
which is aimed at transforming these data from the long to the wide format.
This function produces, among others, the array \code{YY} of response configurations,  
the array \code{XX} of covariate configurations, and the vector \code{freq} of the corresponding frequencies. 

The \proglang{R} function aimed at estimating the class of mixed LM models is 
\code{est\_lm\_mixed}, which requires the following arguments as input:
\begin{itemize}
\item \code{S}: array of response configurations ($n \times T \times r$); 
\item \code{yv}: vector of frequencies of the configurations;
\item \code{k1}: number of support points, corresponding to latent classes, of the distribution of the latent variable $U$;
\item \code{k2}: number of support points, corresponding to latent states, of latent variables denoting the latent process $\b V$;
\item \code{start}: equal to \code{0} for deterministic starting values (default value) and to $1$ for random starting values;
\item \code{tol}: tolerance level for convergence. The default value is \code{10e-8};
\item \code{maxit}: maximum number of iterations before convergence. The default value is \code{1000};
\item  \code{out\_se}: equal to \code{TRUE} to calculate the information matrix and the standard errors. \code{FALSE} is the default option.
\end{itemize}

For the data at hand, the design matrix for the responses, \code{S} (corresponding to \code{YY}), is given by 915 different response  configurations,
$T=6$ age bands, and $r=10$ response variables. Similarly, for the covariate gender, the matrix \code{XX} is given by 915 configurations and $T=6$ age bands. 
In the following we show two different response and covariate configurations and the corresponding
frequency in the sample (given by the vector \code{freq}):

\begin{knitrout}
\definecolor{shadecolor}{rgb}{0.969, 0.969, 
0.969}\color{fgcolor}\begin{kframe}
\begin{alltt}
\hlstd{S} \hlkwb{=} \hlstd{YY}
\hlstd{S[}\hlnum{148}\hlstd{,,]}
\end{alltt}
\begin{verbatim}
##      [,1] [,2] [,3] [,4] [,5] [,6] [,7] [,8] [,9] [,10]
## [1,]    0    0    0    0    0    0    0    0    0     0
## [2,]    0    0    0    0    0    0    0    0    0     0
## [3,]    0    0    0    0    0    0    0    0    0     0
## [4,]    0    0    0    0    0    0    0    0    0     0
## [5,]    0    0    0    0    0    0    0    0    0     0
## [6,]    1    0    0    0    0    0    0    0    0     0
\end{verbatim}
\begin{alltt}
\hlstd{XX[}\hlnum{148}\hlstd{,]}
\end{alltt}
\begin{verbatim}
## [1] 2 2 2 2 2 2
\end{verbatim}
\begin{alltt}
\hlstd{freq[}\hlnum{148}\hlstd{]}
\end{alltt}
\begin{verbatim}
## [1] 3
\end{verbatim}
\begin{alltt}
\hlstd{S[}\hlnum{149}\hlstd{,,]}
\end{alltt}
\begin{verbatim}
##      [,1] [,2] [,3] [,4] [,5] [,6] [,7] [,8] [,9] [,10]
## [1,]    0    0    0    0    1    0    0    0    0     0
## [2,]    0    0    0    0    0    0    0    0    0     0
## [3,]    0    0    0    0    0    0    0    0    0     0
## [4,]    0    0    0    0    0    0    0    0    0     0
## [5,]    0    0    0    0    0    0    0    0    0     0
## [6,]    0    0    0    0    0    0    0    0    0     0
\end{verbatim}
\begin{alltt}
\hlstd{XX[}\hlnum{149}\hlstd{,]}
\end{alltt}
\begin{verbatim}
## [1] 2 2 2 2 2 2
\end{verbatim}
\begin{alltt}
\hlstd{freq[}\hlnum{149}\hlstd{]}
\end{alltt}
\begin{verbatim}
## [1] 113
\end{verbatim}
\end{kframe}
\end{knitrout}

From the above configurations, we observe that only 3 females have been convicted for
violence against the person (first 
response variable)
in the last age band, which is from 36 to 40 years old, 
whereas 113 females committed theft (fifth variable)  during the first time window, related 
to age 10-15. 

To illustrate the use of the function \code{est\_lm\_mixed}, we fit the model in Section \ref{sec_mixed} on such data with $k_1=2$ latent classes,
$k_2=2$  latent states, and we restrict the analysis to females. We use the following command in \proglang{R}:

\begin{knitrout}
\definecolor{shadecolor}{rgb}{0.969, 0.969, 
0.969}\color{fgcolor}\begin{kframe}
\begin{alltt}
\hlstd{S} \hlkwb{=} \hlstd{S[XX[,}\hlnum{1}\hlstd{]}\hlopt{==}\hlnum{2}\hlstd{,,];}
\hlstd{freq} \hlkwb{=} \hlstd{freq[XX[,}\hlnum{1}\hlstd{]}\hlopt{==}\hlnum{2}\hlstd{]}
\end{alltt}
\end{kframe}
\end{knitrout}

\begin{knitrout}
\definecolor{shadecolor}{rgb}{0.969, 0.969, 
0.969}\color{fgcolor}\begin{kframe}
\begin{alltt}
\hlstd{res_mix2f} \hlkwb{=} \hlkwd{est_lm_mixed}\hlstd{(S,} \hlkwc{yv} \hlstd{= freq,} \hlkwc{k1} \hlstd{=} \hlnum{2}\hlstd{,} \hlkwc{k2} \hlstd{=} \hlnum{2}\hlstd{,} \hlkwc{tol} \hlstd{=} \hlnum{10}\hlopt{^-}\hlnum{8}\hlstd{)}
\end{alltt}
\end{kframe}
\end{knitrout}

Then, we get the following convergence info:

\begin{knitrout}
\definecolor{shadecolor}{rgb}{0.969, 0.969, 
0.969}\color{fgcolor}\begin{kframe}
\begin{alltt}
\hlkwd{print}\hlstd{(res_mix2f)}
\end{alltt}
\begin{verbatim}
## Call:
## est_lm_mixed(S = YY, yv = freq, k1 = 2, k2 = 2, tol = 10^-8)
##
## Convergence info:
##      LogLik np   BIC
## [1,] -18347 27 36925
\end{verbatim}
\end{kframe}
\end{knitrout}

The \code{summary} command shows, among others, the estimated mass probabilities $\hat{\lambda}_u$, and the estimated initial and transition probabilities, $\hat{\pi}_{v|u}$ and $\hat{\pi}_{v|u\bar{v}}$:

\begin{knitrout}
\definecolor{shadecolor}{rgb}{0.969, 0.969, 0.969}\color{fgcolor}\begin{kframe}
\begin{alltt}
\hlkwd{summary}\hlstd{(res_mix2f)}
\end{alltt}
\begin{verbatim}
## Call:
## est_lm_mixed(S = YY, yv = freq, k1 = 2, k2 = 2, tol = 10^-8)
## 
## Coefficients:
## 
## Mass probabilities:
## [1] 0.2175 0.7825
## 
## Initial probabilities:
##    u
## v           1       2
##   1 1.000e+00 0.90867
##   2 1.644e-22 0.09133
## 
## Transition probabilities:
## , , u = 1
## 
##    v1
## v0       1      2
##   1 0.8525 0.1475
##   2 0.6414 0.3586
## 
## , , u = 2
## 
##    v1
## v0       1         2
##   1 1.0000 2.455e-15
##   2 0.3382 6.618e-01
## 
## 
## Conditional response probabilities:
## , , j = 1
## 
##    v
## y          1      2
##   0 0.995161 0.8242
##   1 0.004839 0.1758
## 
## , , j = 2
## 
##    v
## y          1      2
##   0 0.998296 0.9809
##   1 0.001704 0.0191
## 
## , , j = 3
## 
##    v
## y          1      2
##   0 0.996297 0.7436
##   1 0.003703 0.2564
## 
## , , j = 4
## 
##    v
## y           1      2
##   0 9.999e-01 0.9737
##   1 9.691e-05 0.0263
## 
## , , j = 5
## 
##    v
## y         1      2
##   0 0.97726 0.4546
##   1 0.02274 0.5454
## 
## , , j = 6
## 
##    v
## y          1      2
##   0 0.998152 0.8892
##   1 0.001848 0.1108
## 
## , , j = 7
## 
##    v
## y          1      2
##   0 0.995658 0.8177
##   1 0.004342 0.1823
## 
## , , j = 8
## 
##    v
## y          1       2
##   0 0.997612 0.91046
##   1 0.002388 0.08954
## 
## , , j = 9
## 
##    v
## y           1       2
##   0 9.999e-01 0.98149
##   1 6.439e-05 0.01851
## 
## , , j = 10
## 
##    v
## y          1      2
##   0 0.998664 0.7912
##   1 0.001336 0.2088
\end{verbatim}
\end{kframe}
\end{knitrout}

The estimated conditional probability of committing each type of crime, $\hat{ \phi}_{j1|v}$,  may be also 
displayed as in the following 

\begin{knitrout}
\definecolor{shadecolor}{rgb}{0.969, 0.969, 0.969}\color{fgcolor}\begin{kframe}
\begin{alltt}
\hlkwd{round}\hlstd{(res_mix2f}\hlopt{$}\hlstd{Psi[}\hlnum{2}\hlstd{,,],}\hlkwc{digits}\hlstd{=}\hlnum{3}\hlstd{)}
\end{alltt}
\begin{verbatim}
##    j
## v       1     2     3     4     5     6     7     8     9    10
##   1 0.005 0.002 0.004 0.000 0.023 0.002 0.004 0.002 0.000 0.001
##   2 0.176 0.019 0.256 0.026 0.545 0.111 0.182 0.090 0.019 0.209
\end{verbatim}
\end{kframe}
\end{knitrout}

According to these estimated probabilities, we can identify the first latent state as that
related to  those females  with null or very low tendency to commit crimes,  whereas the second latent state corresponds to those criminals having mainly as type of activity   
theft, burglary, and some other types of offense. 

The model formulation allows us to characterize the two clusters of individuals at the beginning of the period of observation and to follow their evolution over time. In particular, the first cluster, which includes around 22\% of females, is characterized by subjects that, at the beginning of the period of observation, are all in the first latent state (corresponding to null tendency to commit a crime). On the other hand, females classified in the second cluster (78\%) are characterized by an initial probability of being in the second latent state of around 0.09. 
By comparing the transition probability matrices, we observe, within each cluster, a very high level of persistence in the first latent state. Moreover, females classified in the first cluster present a higher probability (of around 0.64) to move from the second to the first state than those assigned to the second cluster (0.34), revealing a more pronounced tendency to improve in their behavior. 

\section{Conclusions}\label{sec:conclusions}
\vspace{0.5cm}

We propose and illustrate the \proglang{R} package \pkg{LMest} to deal with latent Markov (LM) models.
For a comprehensive overview about these models we refer the reader to \cite{bart:farc:penn:13} and \cite{bart:farc:penn:14}. 
The package allows us to efficiently fit LM models for categorical longitudinal data.
Both the manifest and latent distributions can be parameterized so as to include the effect individual covariates.
The mixed formulation also allows us to include additional latent variables in these parameterizations. 
It shall be noted that all functions above can be used with multivariate categorical
outcomes, with the only exception of \code{est\_lm\_cov\_manifest} which is restricted
to univariate categorical outcomes.

Overall, we consider this package as a relevant advance for applied research in longitudinal data analysis in the presence of categorical response variables.
In particular, we recall that in this context LM models are particularly useful at least from three different perspectives: ({\em i}) to represent and study the evolution of an individual characteristic (e.g., quality of life) that is not directly observable, ({\em ii}) to account for unobserved heterogeneity due to omitted covariates in a time-varying fashion, and ({\em iii}) to account for measurement errors in observing a sequence of categorical response variables.
We recall that, when covariates are available, they are typically included in the measurement model for applications of type ({\em ii}), so that the response variables are affected by observed covariates and latent variables on the same footing, whereas the covariates are included in the latent models for applications of type ({\em i}) and ({\em iii}), so that they affect the distribution of the latent process.

Further updates of the package will include the possibility to use multivariate outcomes in function \code{est\_lm\_cov\_manifest} and new functions with different formulations of mixed LM models, also for sample units collected in clusters \citep{bart:penn:vitt:11}. 
Additionally, the package will allow us to work with continuous outcomes \citep[also
modeling quantiles of these as in][]{farc:10}, and to deal with informative missing values \citep[e.g.,][]{bart:farc:biom:14}. 
We also plan to include estimation methods which are alternative to pure maximum likelihood estimation, as the three-step method proposed by \cite{bart:mont:pand:15}.

\section*{Acknowledgements}
\noindent F. Bartolucci  and F. Pennoni acknowledge the financial support from the grant ``Finite mixture and latent variable models for causal inference and analysis of socio-economic data'' (FIRB - Futuro in ricerca) funded by the Italian Government (RBFR12SHVV).
F. Pennoni also  thanks the financial support of the STAR project ``Statistical models for human perception and evaluation'', University of Naples Federico II. 

\bibliography{biblio}
\bibliographystyle{apalike}

\end{document}